\documentclass[%
 aip,
 amsmath,amssymb,
 reprint,%
]{revtex4-1}

\usepackage{graphicx}
\usepackage{dcolumn}
\usepackage{bm}

\usepackage[english]{babel}
\usepackage[utf8]{inputenc}
\usepackage[T1]{fontenc}
\usepackage{mathptmx}
\usepackage{siunitx}

\usepackage{upgreek}
\usepackage{textcomp} 				
\usepackage{ulem}
\usepackage{color}					
\usepackage{amssymb,amsmath}		

\renewcommand{\paragraph}[1]{{\bf #1}}
\newcommand{\EAFEFE}{$E_{\mathrm{AFE}\rightarrow\mathrm{FE}}$}
\newcommand{\EFEAFE}{$E_{\mathrm{FE}\rightarrow\mathrm{AFE}}$}
\newcommand{\modif}[2]{\textcolor{black}{#2}}

\DeclareUnicodeCharacter{2212}{$--$}

\begin{document}

\preprint{AIP/123-QED}

\title[Critical field anisotropy in the antiferroelectric switching of PbZrO$_3$ films] {Critical field anisotropy in the antiferroelectric switching of PbZrO$_3$ films}

\author{Cosme Milesi-Brault}
 \email{Authors to whom correspondence should be addressed: cosme.milesi@list.lu, mael.guennou@uni.lu}
\author{Nicolas Godard}%
 \affiliation{Materials Research and Technology Department, Luxembourg Institute of Science and Technology, 41 rue du Brill, L-4422 Belvaux, Luxembourg.}
\affiliation{ 
Department of Physics and Materials Science, University of Luxembourg, 41 rue du Brill, L-4422 Belvaux, Luxembourg.
}%
\affiliation{Inter-institutional Research Group Uni.lu--LIST on ferroic materials, 41 rue du Brill, L-4422 Belvaux, Luxembourg}

\author{Stéphanie Girod}
 \affiliation{Materials Research and Technology Department, Luxembourg Institute of Science and Technology, 41 rue du Brill, L-4422 Belvaux, Luxembourg.}
\affiliation{Inter-institutional Research Group Uni.lu--LIST on ferroic materials, 41 rue du Brill, L-4422 Belvaux, Luxembourg}

\author{Yves Fleming}
\author{Brahime El Adib}
\author{Nathalie Valle}
 \affiliation{Materials Research and Technology Department, Luxembourg Institute of Science and Technology, 41 rue du Brill, L-4422 Belvaux, Luxembourg.}

\author{Sebastjan Glin\v sek}
\author{Emmanuel Defay}
 \affiliation{Materials Research and Technology Department, Luxembourg Institute of Science and Technology, 41 rue du Brill, L-4422 Belvaux, Luxembourg.}
\affiliation{Inter-institutional Research Group Uni.lu--LIST on ferroic materials, 41 rue du Brill, L-4422 Belvaux, Luxembourg}

\author{Mael Guennou}
 \email{Authors to whom correspondence should be addressed: cosme.milesi@list.lu, mael.guennou@uni.lu}
\affiliation{ 
Department of Physics and Materials Science, University of Luxembourg, 41 rue du Brill, L-4422 Belvaux, Luxembourg.
}%
\affiliation{Inter-institutional Research Group Uni.lu--LIST on ferroic materials, 41 rue du Brill, L-4422 Belvaux, Luxembourg}

\begin{abstract}
Antiferroelectrics have been recently sparking interest due to their potential use in energy storage and electrocaloric cooling. Their main distinctive feature is antiferroelectric switching, i.e.\ the possibility to induce a phase transition to a polar phase by an electric field. Here we investigate the switching behavior of the model antiferroelectric perovskite PbZrO$_3$ using thin films processed by chemical solution deposition in different geometries and orientations. Both out-of-plane and in-plane switching configurations are investigated. \textcolor{black}{The critical field is observed to be highly dependent on the direction of the electric field with respect to the film texture. We show that this behaviour is qualitatively consistent with a phase transition to a rhombohedral polar phase.} 
We finally estimate the importance of crystallite orientation and film texturation in the variations observed in the literature.
\end{abstract}

\maketitle


Antiferroelectric (AFE) materials are recognized by a set of experimental signatures: a phase transition between two non-polar phases exhibiting an anomaly of the dielectric constant, antiparallel atomic displacements identified as sublattices of electric dipoles, and finally, in their low-symmetry phase, \modif{a characteristic double hysteresis loop of the polarization as a function of electric field $P(E)$, which reflects the presence of a polar polymorph close in energy to the non-polar ground state. This double loop}{the possibility to induce a phase transition to a polar phase by an electric field~\cite{Rabe2013, Toledano2016}. This electric-field-induced phase transition between an antipolar and a polar state, here called "antiferroelectric switching",} is probably the most remarkable signature of AFE materials, and \modif{this antiferroelectric switching}{} the property that is most promising for a practical use of antiferroelectric materials, as exemplified by capacitors for energy storage \cite{Liu2018} or antiferroelectric tunnel junctions~\cite{Apachitei2017}. Yet, it has been studied in details in a very limited number of cases. 

The model antiferroelectric perovskite, PbZrO$_3$, is in that respect particularly complex. Its AFE transition involves complex couplings and several order parameters, including oxygen octahedra tilts \cite{Tagantsev2013,Iniguez2014, Xu2019}. In the AFE phase, lead displacements are considered to be along the orthorhombic $a$ axis (i.e.\ along a [110]-pseudo cubic direction). The transition to the polar phase, however, probably does not proceed through a simple flipping of a sublattice but by a transition to a totally different phase with a different pattern of octahedra tilts. It is usually admitted that the polar phase of PbZrO$_3$ is rhombohedral with polarization along a [111]$_\mathrm{pc}$ direction; as inferred from the presence of a rhombohedral phase in the phase diagram of PZT at low Ti concentrations~\cite{JAFFE1971115}. First-principle calculations have confirmed that the energy of the polar rhombohedral phase is indeed very close to the antiferroelectric polymorph \cite{Singh1995,Kagimura2008, Reyes-Lillo2013, Gao2017}, but other structures have been proposed~\cite{Leyderman1998,Toledano2019, Lisenkov2020}. Experimentally, early diffraction, optical and dielectric studies seemed to confirm the rhombohedral hypothesis~\cite{Shirane1954,Fesenko1976,Fesenko1978,Fesenko1985} \textcolor{black}{but there are very little direct structural studies under electric field, and no real consensus on the symmetry of the polar phase}. 

This relative lack of knowledge, even on materials considered as models, hinders the optimization of antiferroelectric switching. The difficulty largely finds its origin in the very high electric fields needed to switch bulk samples at ambient conditions. Thin films or multilayer capacitors, in contrast, can usually sustain much higher fields before breakdown~\cite{Wang1992}. In this work, we use this approach and investigate the importance of film texture, i.e. the crystallographic orientation of the grains, on the switching properties of antiferroelectric PbZrO$_3$ films. We demonstrate both out-of-plane and in-plane switching, using two different sample geometries. We compare their characteristics within a simple switching model, and estimate the importance of this parameter based on a review from literature data. 


\begin{figure*}[htpb]
    \centering
    \includegraphics[width=\textwidth]{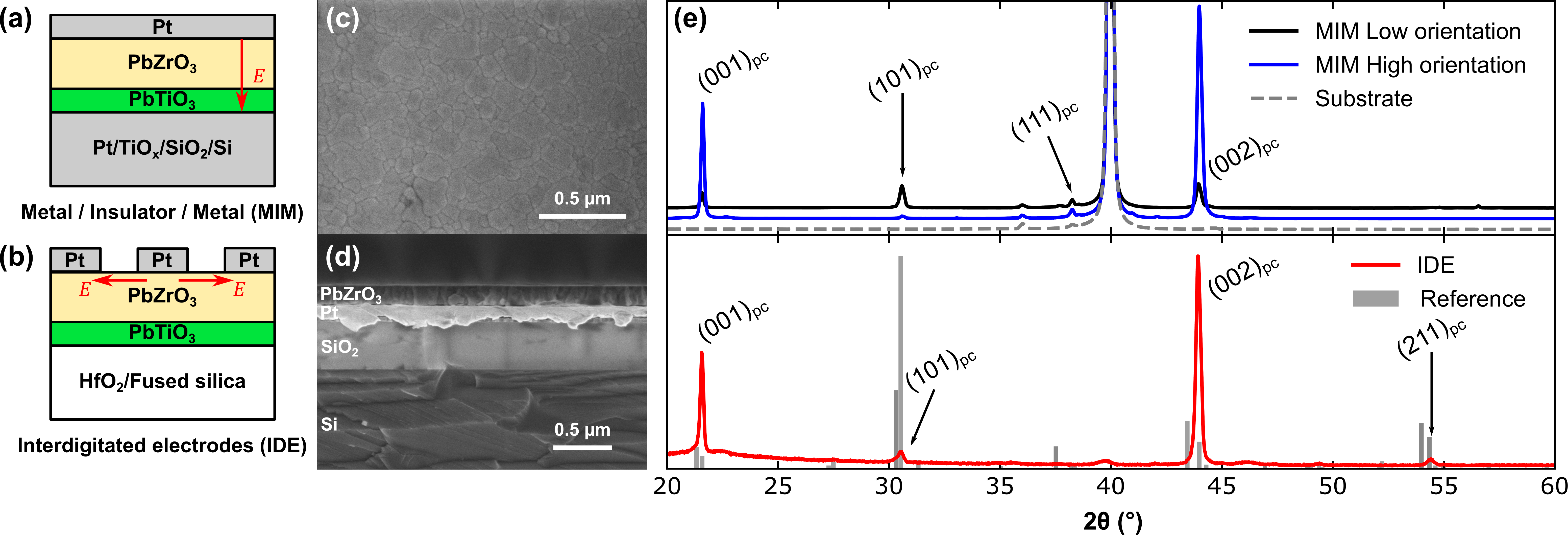}
    \caption{Representation of the different sample geometries: (a) Metal/Insulator/Metal (MIM) geometry and (b) Interdigitated electrodes (IDE) geometry. Representative SEM micrographs of (c) the surface and (d) a cross-section of a PbZrO$_3$ film in a MIM geometry. (e) XRD patterns of 255 nm-thick PbZrO$_3$ films in both MIM (top) and IDE (bottom) geometries.
    }
    \label{fig:geometries-XRD-SEM}
\end{figure*}

Lead zirconate (PbZrO$_3$) films were deposited by chemical solution deposition (CSD) on two different substrates: platinized silicon (Si \SI{675}{\micro \metre}/SiO$_2$ \SI{500}{\nm}/TiO$_x$ \SI{20}{\nm}/Pt \SI{100}{\nm}) and fused silica covered with a 23 nm-thick ALD-deposited HfO$_2$ buffer layer. In both cases, \modif{seeds}{a seed layer} of lead titanate (PbTiO$_3$) were deposited prior to PbZrO$_3$. Lead zirconate solutions were processed following a standard process commonly used for PZT \cite{Godard2019} (details in the supplementary information). Films with thicknesses of \SI{85}{\nm}, \SI{170}{\nm} and \SI{255}{\nm} were obtained. On platinized silicon, Pt electrodes were sputtered to obtain a Metal-Insulator-Metal (MIM) geometry, as sketched in Fig.~\ref{fig:geometries-XRD-SEM}.a. On fused silica, Pt interdigitated electrodes (IDE) with a gap of \SI{5}{\um} were sputtered and patterned with lift-off photolithography (Fig.~\ref{fig:geometries-XRD-SEM}.b.). Surface and cross-section imaging of the films was done by a Helios NanoLab scanning electron microscope from FEI; the surface appears to be non-porous and defect-free (Fig.~\ref{fig:geometries-XRD-SEM}.c.), and the cross-sections show dense films of homogeneous thickness with a columnar structure \textcolor{black}{and lateral grain sizes ranging from 85 to \SI{115}{\nm}} (Fig.~\ref{fig:geometries-XRD-SEM}.d. and Supplemental Information). \textcolor{black}{SIMS depth-profile analyses have been performed showing that the Pb/Zr ratio through the film is similar in both geometries, and a limited diffusion of Ti into the PbZrO$_3$ in the MIM samples (Supplemental Information).}

X-ray diffraction measurements were performed on a Bruker D8 Discover diffractometer equipped with a Goebel mirror and a 5-axis cradle in the Bragg-Brentano geometry ($\lambda_{\mathrm{Cu} (K_\alpha)} =$~\SI{1.54184}{\angstrom}) and confirmed that the films crystallized in the perovskite phase, without any indication of parasitic phase (Fig.~\ref{fig:geometries-XRD-SEM}.\textcolor{black}{e.}). Besides, all films show a pronounced preferred orientation along a [001]$_\mathrm{pc}$ direction  as shown in Fig.~\ref{fig:geometries-XRD-SEM}.e. Comparing with a reference powder diffraction pattern \cite{Fujishita1998} in the orthorhombic AFE phase ($Pbam$ space group), we conclude that the films are oriented along the [002]$_\mathrm{o}$ direction. Pole figures confirmed that the orientation is otherwise isotropic in the plane of the films (Supplemental Information). For films on Pt/Si, the different processes for the PbTiO$_3$ \modif{seeds}{seed layer} resulted in two different degrees of preferred orientation, which we quantified using Lotgering factors~\cite{Lotgering1959}. The films with the highest preferred orientation in the MIM and IDE geometries have Lotgering factors $f_{00l}$ around 0.96 and 0.84 respectively, i.e.\ are equally well oriented. The MIM sample with a lower degree of preferred orientation has a factor $f_{00l} \approx$ 0.50 only.

For the studies of the antiferroelectric switching loops, a TF Analyzer 2000 from aixACCT was used to record the polarization-electric field loops by sending a bipolar triangular voltage ramp with a frequency of \SI{100}{\Hz}. The frequency dependence of the different \textcolor{black}{critical fields} has been studied between \SI{10}{\Hz} and \SI{5}{\kHz}, and was found to have little influence on the results, as shown in the supplementary information. In the following, the fields required to switch from the antiferroelectric to the ferroelectric state and back are noted \EAFEFE{} and \EFEAFE, respectively. 

\begin{figure*}[htpb]
    \centering
    \includegraphics[trim={0 0 0 0}, clip, width=\textwidth]{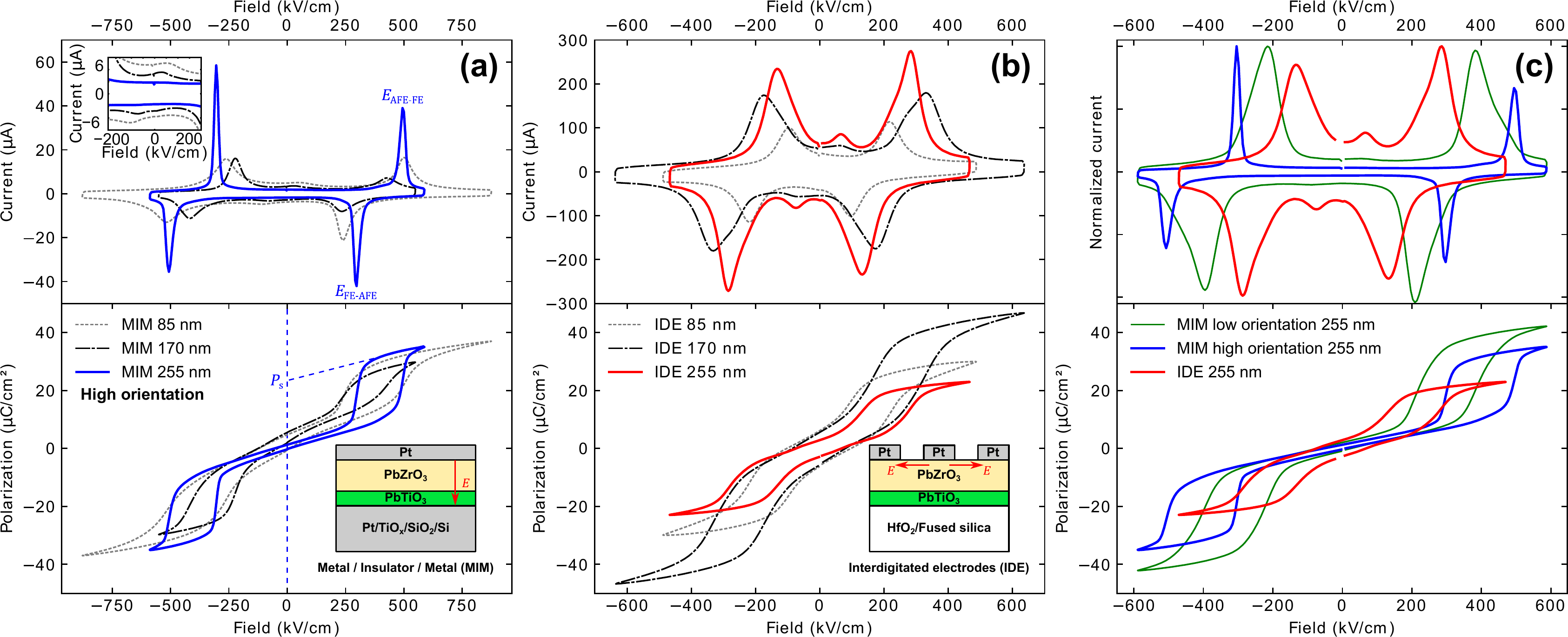}
    \caption{Polarization and current curves for (a) highly oriented MIM PZO films (thickness between 85 and 255 nm), (b) IDE geometry (thickness between 85 and 255 nm). (c) shows a comparison between MIM low orientation, MIM high orientation and IDE 255 nm-thick PZO films.
    In insert in (a) is a zoom on the current curves in the $-200$ to \SI{200}{\kV\per\cm} region to highlight the presence of a "ferroelectric-like" peak.}
    \label{fig:all_aixacct}
\end{figure*}

The AFE switching of the films with the MIM geometry (out-of-plane switching) is shown in Fig.~\ref{fig:all_aixacct}.a. Here, the electric field is simply defined as the applied voltage divided by the film thickness. The double hysteresis loop is seen in all cases, but the \textcolor{black}{critical fields} show some variations, and current switching peaks become thinner as film thickness increases. The antiferroelectric switching loops for samples with IDE (in-plane switching) are shown in Fig.~\ref{fig:all_aixacct}.b. Here, in order to derive an electric field from the applied voltage, we followed the procedure described in Ref.~\cite{Nigon2017} and calculated the electric field as $E=V/(a+\Delta a)$ where $a$ is the gap between IDE fingers and $\Delta a$ depends only on the film thickness $t_f$ as $\Delta a \approx 1.324~t_f$. 
Qualitatively, the double loops for the different sample thickness show similar trends. In particular, the loop for the thinner films is significantly more opened at zero field. We compare in Fig.~\ref{fig:all_aixacct}.c. the double hysteresis loops in both MIM and IDE geometries for 255 nm-thick samples. The comparison shows that the switching in the IDE sample occurs at a lower electric field, and with a significantly broader current peak. We also include in this comparison the sample with a lower preferred orientation. It is markedly different from the other MIM samples, with broader current peaks at lower electric fields. The difference in critical field observed between MIM and IDE samples clearly exceeds the variations seen with sample thickness. \textcolor{black}{It also largely exceeds the small variations observed on a given sample (less than 2\%  for highly-oriented MIM and 5\% for IDE samples, as checked on 3 to 5 different locations).} We attribute the difference primarily to the different \textcolor{black}{orientation of the electric field with respect to the} texture of the films. Overall, the sample with the best orientation in the MIM geometry is the one displaying the sharpest switching behaviour, with thin current peaks and a correspondingly sharp hysteresis.

In addition to the main antiferroelectric switching signal, we observe in the $I(V)$ curves a smaller but significant current peak at lower electric fields, typically below \SI{100}{\kV\per\cm}. This peak is particularly noticeable in the IDE geometry (Fig.~\ref{fig:all_aixacct}.b.), but is also present in the MIM geometry for the thinner films (inset in Fig.~\ref{fig:all_aixacct}.a.). The presence of additional ferroelectric displacements of lead ions is not unheard of in PbZrO$_3$ in the context of structural studies~\cite{Sawaguchi1951, Jona1954, PASTO1973, Dai1995}, and similar features have even been reported in the switching loops of PbZrO$_3$ thin films~\cite{Pintilie2008}. However, in our case, this contribution could also originate from the presence of the additional PbTiO$_3$ \modif{seeds}{seed layer} and we cannot make a conclusive statement about its origin. 

\begin{figure}[htpb]
    \centering
    \includegraphics[trim={0 0 0 0}, clip, width=0.48\textwidth]{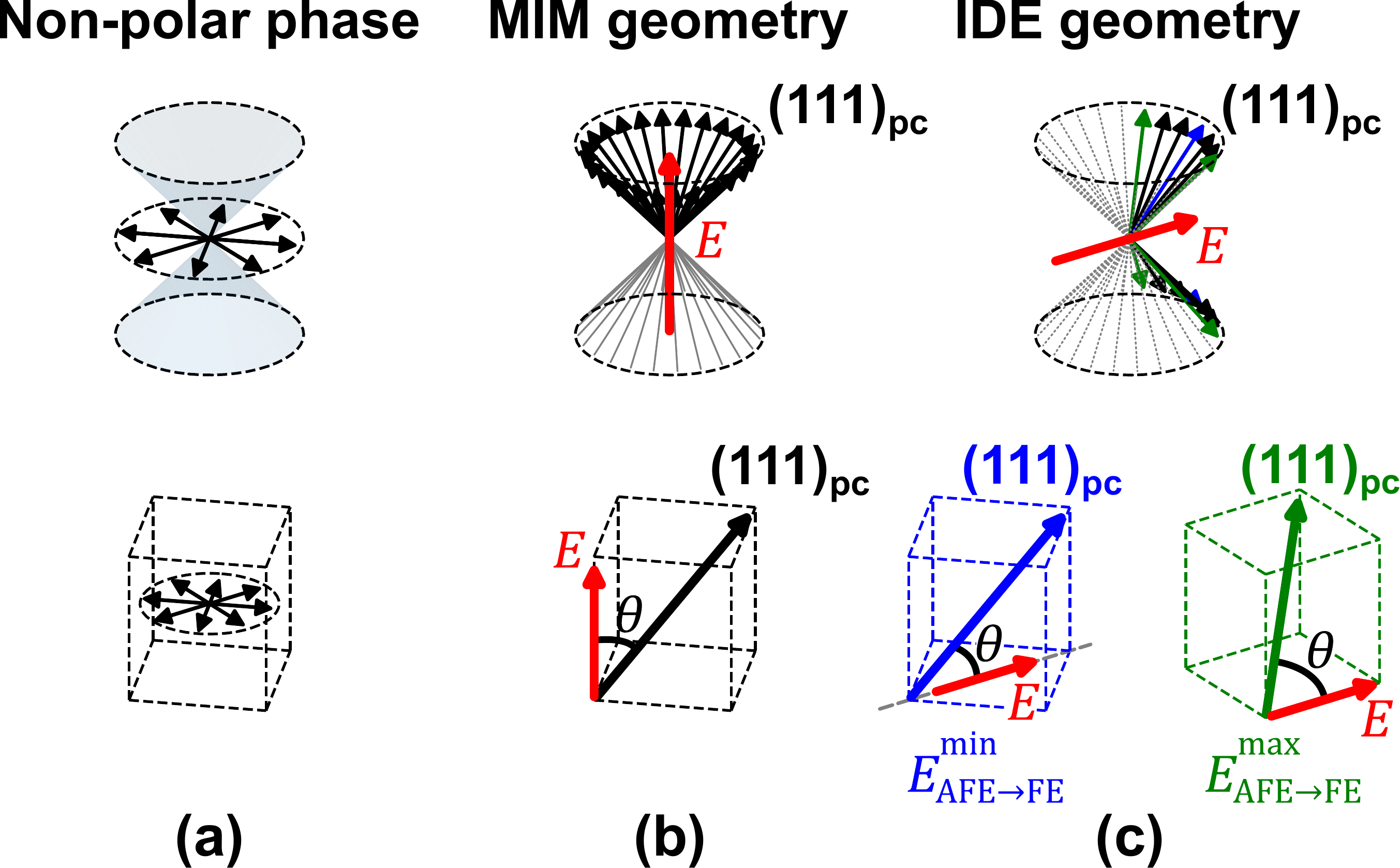}
    \caption{Sketch of the dipole arrangements expected in the non-polar and polar phases in our PbZrO$_3$ films, assuming a perfect preferred orientation and a rhombohedral polar phase.}
    \label{fig:switching}
\end{figure}

We now discuss the characteristics of the AFE switching loops in relation to the film textures determined by XRD. We will assume that the polar phase of PbZrO$_3$ is rhombohedral with a polarization along a (111)$_\mathrm{pc}$-direction, and that the switching occurs when the energy of the polar phase under field reaches some threshold. For a given crystallite, the \textcolor{black}{critical field} is expected to be lowest if the field is applied along a (111)$_\mathrm{pc}$-direction, highest for a field applied along (001)$_\mathrm{pc}$-direction, and to otherwise scale with $1/\cos\theta$ where $\theta$ is the angle between the applied electric field and the closest (111)$_\mathrm{pc}$-direction. For a polycrystalline film, switching will occur according to a distribution of switching events and fields determined by the texture of the film.  

This crude toy model is qualitatively consistent with the observations. With a perfect texture, i.e.\ all grains oriented along a (001)$_\mathrm{pc}$ direction, the \textcolor{black}{critical field} in the MIM geometry is the same for all grains, therefore the sharp current peak, and with the highest possible value (Fig~\ref{fig:switching}.b). In contrast, the switching peaks for the film with a lower degree of preferred orientation are found to be broader and at lower field values. 
\textcolor{black}{When the field is applied in-plane with IDEs, for a given crystallite, the electric field can lie in any direction in the (001)$_\mathrm{pc}$ plane, and we expect the film as a whole to switch according to a distribution of critical fields. The lowest and highest possible critical fields are obtained when it is aligned along a (110)$_\mathrm{pc}$ direction and a (100)$_\mathrm{pc}$ direction, respectively, of the crystallite, as illustrated in Fig.~\ref{fig:switching}.(c).}
The former happens to be equal to the field in the MIM geometry; the latter is expected to be lower by a factor of 1.41. These field values are qualitatively consistent with Fig.~\ref{fig:all_aixacct}.c. Indeed, in the MIM geometry, the switching current peaks are much sharper (FWHM $\approx$ \SI{30}{\kV\per\cm}) than in the IDE case (FWHM $\approx$ \SI{100}{\kV\per\cm}) and also shifted towards lower electric field values. This trend is verified for all sample thicknesses, regardless of the variations between samples of a given geometry. \textcolor{black}{Therefore, the value of the critical fields appear to be determined predominantly by the orientation of the field with respect to the crystallographic axes of PbZrO$_3$.}

Investigations into the literature supports the assumption that the small variations between samples described in the supporting information have in comparison a minor influence of the critical field. Grain size and grain boundaries are in particular potentially important factors. The grain sizes in our samples are not in a range where strong size effects are expected~\cite{Chattopadhyay1997}, and their limited distribution (\SIrange[range-phrase = --]{85}{115}{\nm}) is not expected to cause large variations in critical field~\cite{Kong2002}. The same can be said about the residual strains resulting from differences in thermal expansion between film and substrate: \textcolor{black}{both MIM and IDE samples are in a tensile strain state with different magnitudes, but comparison with previous works showing wider variations in strain and thicknesses hints that this should not be determinant~\cite{Ge2013,Corkovic2008}.} The anisotropy of the dielectric constant might conceivably play a role, but was found to be weak in Ref.~\onlinecite{Kanno2006}. Our own capacitance measurements on two 255 nm-thick MIM samples with different degrees of orientation (high and low) also did not reveal any significant difference (Supplemental Information), in spite of their very different switching characteristics.
All this supports that the direction of the electric field with respect to the preferred orientation of the film is the main parameter influencing the critical field in our samples.

\textcolor{black}{Applying the electric field normal to or within the sample plane does not only make a difference with respect to the crystallographic axes of PbZrO$_3$, but also with respect to the orientation of the grain boundaries separating the columnar grains. In ferroelectrics, it has been suggested that grain boundaries can potentially influence the critical field in both ways: lower it if grain boundaries act as nucleation points\cite{Meier2016}, or increase it if they cause an additional potential drop\cite{Amorin2010}. Here, considering that the critical field in IDE samples is even lower than what we would expect from the reasoning on crystallographic orientation alone, we hypothesize that grain boundaries perpendicular to the field facilitate the in-plane switching process, which could be related to the nucleation scenario, or a better accommodation of the strain accompanying the AFE-FE transition.}


\begin{figure}[htpb]
    \centering
    \includegraphics[trim={0 0 0 0}, clip, width=0.48\textwidth]{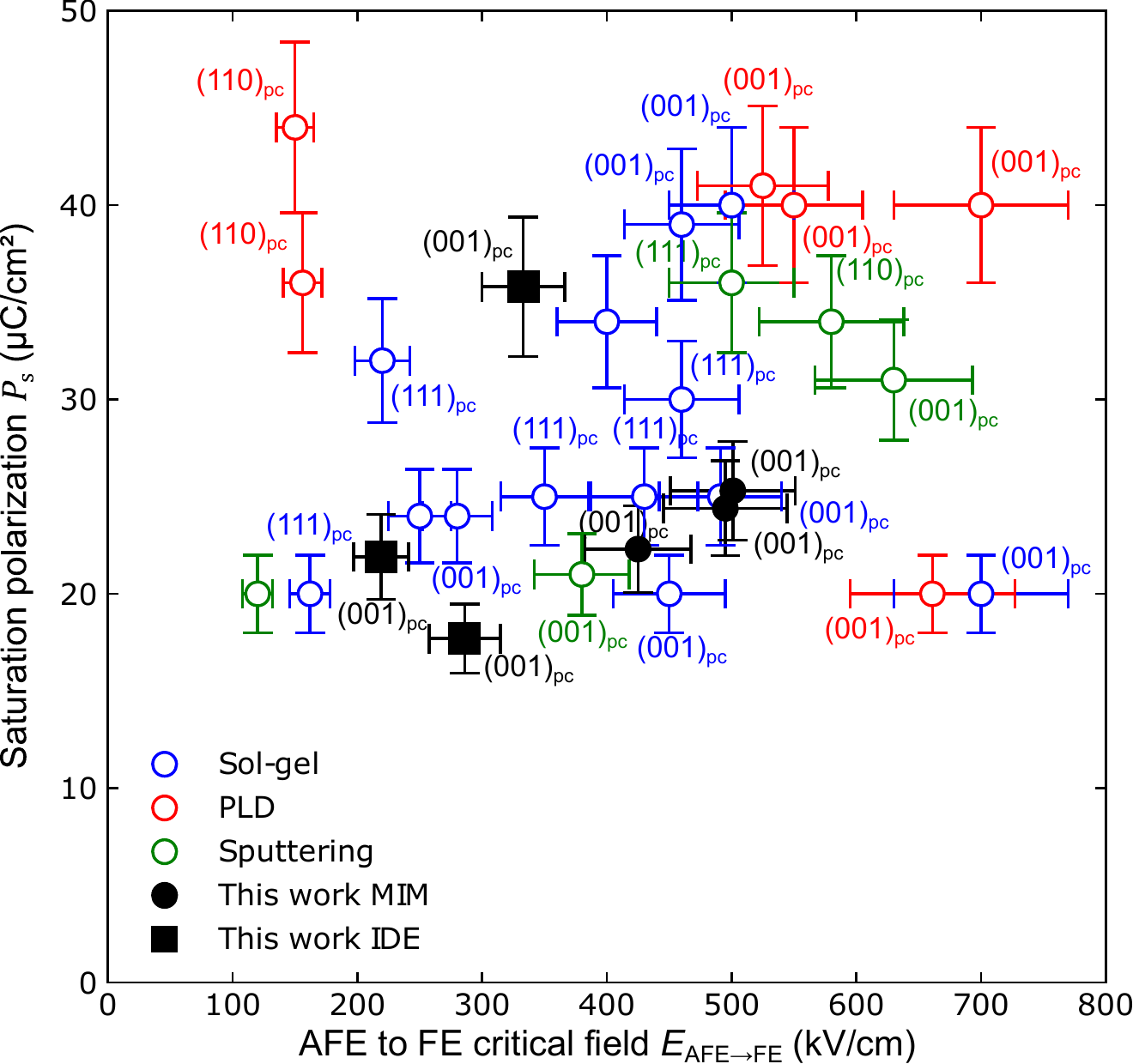}
    \caption{Comparison between our data and sol-gel/PLD/sputtering samples from literature of \textcolor{black}{critical fields} \EAFEFE{} and saturation polarization $P_\mathrm{s}$. Details about orientations and other aspects are given in the supplementary information.}
    \label{fig:litterature}
\end{figure}

In order to check whether this trend and the influence of \textcolor{black}{film orientation} is confirmed by literature data, we reviewed the papers reporting on the AFE switching at room temperature of PbZrO$_3$ films processed by sol-gel and other  techniques~\cite{Tang2004,Hao2011,Nguyen2020,Hao2009a,Tang2003,Zhai2003,Guo2019,Ye2011,Zhai2002,Wang1992,Li1995,Bharadwaja2000,Bharadwaja1999,Dobal2001,Boldyreva2007,Nguyen2018a,Guo2017,Ge2014,Kanno2006}. We chose to focus on the values for \EAFEFE{} and $P_\mathrm{s}$, which are more commonly reported, and defined $P_\mathrm{s}$ as the saturation polarization as depicted in Fig.~\ref{fig:all_aixacct}.a. There is inevitably some uncertainty in the process of defining, reporting and reading those data, and generous error bars have been estimated to be $\pm 10 \%$ of the \EAFEFE{} and $P_\mathrm{s}$ values. All data are graphically summarized in Fig.~\ref{fig:litterature}, and the full list, with the details of references and additional data, is given in the Supporting Information. Fig.~\ref{fig:litterature} reveals some clear trends. Most importantly, for sol-gel processed films, we notice that the trend identified in our data is indeed confirmed: (111)$_\mathrm{pc}$-oriented samples form a group with much lower critical fields than the group of (001)$_\mathrm{pc}$-oriented samples, on the average, regardless of the differences that these films from different sources are bound to exhibit. This points to the importance of  \textcolor{black}{film orientation} as an essential way to minimize -- or control in general -- the critical field required for antiferroelectric switching.  

On the other hand, Fig.~\ref{fig:litterature} also shows the limits of a simple geometric reasoning. The critical field values are clearly scattered over a rather broad distribution, and the agreement with the switching model is only qualitative, as seen in the ratios between critical fields. \textcolor{black}{This must be attributed to the other factors affecting the values of the critical fields. It is difficult at this stage to investigate possible correlations and identify dominant factors, due to the unequal level of characterization details reported in this body of literature. This will require more systematic work. Interestingly}, we note nonetheless that films grown by sputtering and PLD give trends that are very different from the trends in polycrystalline films, e.g. with lower critical fields observed for (001)$_\mathrm{pc}$-oriented films in Ref.~\cite{Ge2014}, suggesting that the strong epitaxial strains plays in that case a dominant role.

Finally, let us point out that the rhombohedral structure for the polar phase of PbZrO$_3$ was here hypothesized, and found consistent with the observations. In contrast, the same reasoning with the assumption of a tetragonal, or tetragonal-like polar structure, i.e.\ with a polarization aligned with or close to a $(001)_\mathrm{pc}$ direction, would be completely inconsistent with the observations, even at a qualitative level: the lowest fields in our films would be expected in the MIM geometry for the fully oriented films, which is clearly not the case. This gives a strong argument to discard the hypothesis of a tetragonal-like polar phase of PbZrO$_3$. More generally, in the context of antiferroelectrics where switching of oriented single crystals may be difficult to achieve, \textcolor{black}{electrical} studies of \textcolor{black}{oriented} films appear as a promising way to address this structural question.  

In summary, we have presented the synthesis and characterization of oriented polycrystalline antiferroelectric PbZrO$_3$ films. X-ray diffraction shows highly-oriented and more randomly oriented PbZrO$_3$ thin films and electrical measurements confirm that a sharpest transition is directly linked to better orientation. AFE switching is demonstrated in a standard MIM geometry, but also more originally by applying the electric field in the plane of the films via IDEs. By comparison with the available data in the literature, we show that film or ceramic \textcolor{black}{orientation} can be a major way to control the \textcolor{black}{critical field}. Besides, our results are consistent with the generally accepted hypothesis of a rhombohedral, or rhombohedral-like, structure of the field-induced phase of PbZrO$_3$. \textcolor{black}{Studying the antiferroelectric switching of oriented films along different directions therefore appears as a possible approach to investigate the structure of field-induced polar phases in antiferroelectric materials in general.} 

\section*{Supplementary Material}

See supplementary material for more detailed parameters used in the Chemical Solution Deposition (CSD) synthesis, Scanning Electron Microscopy (SEM) surface and cross-section micrographs, Secondary Ion Mass Spectrometry (SIMS) analysis, XRD pole figures, XRD strain measurements, frequency dependence of polarization and capacitance measurements and a table summarizing literature data.

\section*{Acknowledgments}

The authors would like to thank Veronika Kovacova and Tony Schenk for their help in the XRD strain measurements. The authors acknowledge financial support by the Luxembourg National Research Fund under project BIAFET C16/MS/11348912/Guennou and CAMELHEAT C17/MS/11703691/Defay. 

Processing details disclosed in the present contribution are protected by a patent filed in June 2020 (N. Godard, S. Glin\v sek, A. Bl\'azquez Mart\'inez, E. Defay, patent LU101884, 2020).

\section*{Data availability}

The data that support the findings of this study are available from the corresponding author upon reasonable request.

\newpage

\bibliography{PZO}

\end{document}